\documentstyle[12pt]{article} 
\newcommand{\be}{\begin{equation}}
\newcommand{\ee}{\end{equation}}
\newcommand{\bea}{\begin{eqnarray} \nonumber}
\newcommand{\eea}{\end{eqnarray}}
\newcommand{\beaa}{\begin{eqnarray}}
\newcommand{\eeaa}{\end{eqnarray}}
\newcommand{\ba}{\begin{array}}
\newcommand{\ea}{\end{array}}
\def\lab{\label}

\def\le{\left}
\def\ri{\right}

\newcommand{\bit}{\begin{itemize}}
\newcommand{\eit}{\end{itemize}}
\newcommand{\ben}{\begin{enumerate}}
\newcommand{\een}{\end{enumerate}}

\def\sma#1{\mbox{\footnotesize $#1$\normalsize}}

\begin{document} 

\centerline{{\Large \bf Phase Coherence in Quantum Brownian Motion}}

\vspace{1.7cm}

\centerline{\large 
M. Blasone\footnote{blasone@vaxsa.csied.unisa.it}, 
Y.N. Srivastava\footnote{srivastava@pg.infn.it}, 
G. Vitiello\footnote{vitiello@vaxsa.csied.unisa.it} 
and A. Widom\footnote{widom@neu.edu}}

\vspace{0.7cm}

\centerline{${}^{1,3}$Dipartimento di Fisica, Universit\`a di Salerno, 
84100 Salerno, Italy}
\centerline{and Gruppo Collegato INFN Salerno, Sezione di Napoli}

\vspace{0.2cm}

\centerline{${}^{2,4}$Physics Department, Northeastern University, 
Boston MA., U.S.A.}
\centerline{Dipartimento di Fisica, INFN, 
Universit\`a di Perugia, Perugia, Italy}

\vskip .7in

\centerline{\bf Abstract}

\bigskip

\small

The quantum theory of Brownian motion is discussed in the Schwinger 
version wherein the notion of a coordinate moving forward in time 
$x(t)$ is replaced by two coordinates, $x_+(t)$ moving forward in time 
and $x_-(t)$ moving backward in time. The role of the doubling of 
the degrees of freedom is illustrated for the case of electron beam 
two slit diffraction experiments. Interference is computed with 
and without dissipation (described by a thermal bath). The notion of 
a dissipative interference phase, closely analogous to the Aharonov-Bohm 
magnetic field induced phase, is explored.


\normalsize

\newpage

\section{Introduction}

In the Bohr Copenhagen interpretation of quantum measurements, data are 
taken from a classical apparatus reading which is influenced by a quantum 
object during the time interval in which they interact. Most of the 
theoretical work in analyzing quantum measurements requires computations 
for the quantum object. However, Bohr's dictate is that {\it one must not 
ask what the quantum object is really doing}! All that can be said is 
that the classical apparatus determines which of the complementary aspects 
of the quantum object will be made manifest in the experimental data.

As applied to the two slit particle diffraction experiment \cite{TO}, what 
this means, dear reader, is that you will never know how a single particle 
managed to have non-local awareness of two slits. Furthermore, you are not 
even allowed to ask the question because it cannot be answered 
experimentally without destroying the quantum interference diffraction 
pattern.

But the question was asked repeatedly in different forms by Einstein, 
who insisted that the picture is incomplete until such time that we can 
assign some objective reality as to what the quantum object is doing. The 
question does not merely concern the fact that  nature plays a game of 
chance and forces us to use probabilities. In his work on Brownian motion, 
Einstein derived a relation between the Brownian particle diffusion 
coefficient $D$ and the mechanical fluid induced friction 
coefficient $R$, 
\be
D={k_BT\over R}, \lab{(1)}
\ee 
which allowed experimentalists to verify both the existence of atoms 
(which many physicists had previously doubted) and the correctness of 
the statistical thermodynamics of Boltzmann and Gibbs. We may safely 
assume that Einstein knew that nature played games of chance. But a 
Brownian particle in a fluid does something real. It jumps randomly 
back and forth and if it is large enough you are allowed to observe 
this motion in some detail. 

What picture can we paint for the motion of a quantum Brownian particle? 
We will ignore Bohr's injunction about never being able to know and try 
to form a picture. The formalism for dealing with quantum Brownian motion 
was developed in complete generality by Schwinger and presented to the 
physics community with a pleasant sense of humor \cite{SW}. 
None of the Schwinger's 
many lengthy equations are numbered, and the central result concerning 
general quantum Brownian motion in the presence of non-linear 
forces was quoted without any derivation at all. Nevertheless, Schwinger's 
formalism is mathematically complete and the results will be used by us 
for simple quantum Brownian motion consistent with the Einstein 
Eq.(\ref{(1)}). 

The physical picture of quantum Brownian motion has two parts: 

(i) The 
starting point is that a classical object can be viewed as having (say) 
a coordinate which depends on time $x(t)$. A quantum object may be viewed 
as splitting the single coordinate $x(t)$ into two coordinates $x_+(t)$ 
(going forward in time) and $x_-(t)$ (going backward in time) \cite{SW}. 
The classical 
limit is obtained when both motions coincide $x(t)=x_+(t)=x_-(t)$. To see 
why this is the case, one may employ the Schwinger quantum operator action 
principle, or more simply recall the mean value of a quantum operator 
\be
\bar{A}(t)=(\psi (t)|A|\psi (t))=
\int \! \int  \psi^* (x_-,t)\, (x_-|A|x_+)\,\psi (x_+,t)\;dx_+ dx_-. 
\lab{(2)}
\ee
Thus one requires two copies of the Schr\"odinger equation to follow the 
density matrix
\be
(x_+|\rho (t)|x_-)=\psi^* (x_-,t)\psi (x_+,t), \lab{(3)}
\ee
i.e. the forward in time motion
\be
i\hbar {\partial \psi (x_+,t) \over \partial t}=H_+\psi (x_+,t), \lab{(4a)}
\ee
and the backward in time motion
\be
-i\hbar {\partial \psi^* (x_-,t) \over \partial t}=H_-\psi^* (x_-,t), 
\lab{(4b)}
\ee
yielding 
\be
i\hbar {\partial (x_+|\rho (t)|x_-) \over \partial t}=
{\cal H}\ (x_+|\rho (t)|x_-), \lab{(5a)}
\ee
where  
\be
{\cal H}=H_+ -H_-. \lab{(5b)}
\ee

The requirement of working with two copies of the Hamiltonian 
(i.e. $H_{\pm }$) operating on the outer product of two Hilbert 
spaces has been implicit in quantum mechanics since the very beginning 
of the theory. For  example, from Eqs.(\ref{(5a)}), (\ref{(5b)}) 
one finds immediately that 
the eigenvalues of the dynamic operator ${\cal H}$ are 
directly the Bohr transition frequencies $\hbar \omega_{nm}=E_n-E_m$ 
which was the first clue to the explanation of spectroscopic structure. 

If one accepts the notion of both forward in time and backward in time 
Hilbert spaces, then the following physical picture of two slit 
diffraction emerges. The particle can go forward and backward in time 
through slit 1. This is a classical process. The particle can go forward 
in time and backward in time through slit 2, which is also 
classical since for classical cases $x_+(t)=x_-(t)$. On the other hand, 
the particle can go forward in time through slit 1 and backward in time 
through slit 2, or forward in time through slit 2 and backward in time 
through slit 1. These are the source for quantum 
interference since $|x_+(t)-x_-(t)|>0$. The notion that a quantum particle 
has two coordinates $x_{\pm }(t)$ moving at the same time is central. 
In Sec.2 we show by explicit calculation of diffraction patterns that it 
is the difference between the two motions 
\be
y=x_+ -x_- \lab{(6)}
\ee
that induces quantum interference. 

(ii) The second part of the picture involves the question of how a 
classical situation with $x_+(t)=x_-(t)$ arises. In Sec.3, the Brownian 
motion of a quantum particle is discussed along with the damped evolution 
operator modification of Eqs.(\ref{(5a)}), (\ref{(5b)})
\cite{DO} which becomes 
(for a Brownian particle of 
mass $M$ moving in a potential U(x) with a damping resistance R) 
\cite{SWV, CS, TS}
\be
{\cal H}_{Brownian}= \frac{1}{2M}
\le(p_+ - \frac{R}{2}\,  x_-\ri)^2-\frac{1}{2M}
\le(p_- + \frac{R}{2}\,  x_+\ri)^2
+U(x_+)-U(x_-)-
{ik_BTR \over \hbar}(x_+-x_-)^2, \lab{(7a)}
\ee
\be
p_\pm =-i\hbar {\partial \over \partial x_\pm}, \lab{(7b)}
\ee
\be
i\hbar {\partial (x_+|\rho (t)|x_-) \over \partial t}=
{\cal H}_{Brownian}\ (x_+|\rho (t)|x_-), \lab{(7c)}
\ee
where the density operator in general describes a mixed statistical state. 
In Sec.3 it will also be shown that the thermal bath contribution to 
the right hand side of Eq.(\ref{(7a)}) 
(proportional to fluid temperature T) is 
equivalent to a white noise fluctuation source coupling the forward and 
backward motions in Eq.(\ref{(6)}) according to
\be
<y(t)y(t^\prime )>_{noise}={\hbar^2 \over 2Rk_BT}\delta (t-t^\prime ), 
\lab{(8)}
\ee 
so that continual thermal fluctuations are always occurring in the 
difference Eq.(\ref{(6)}) between forward in time and backward in time
coordinates. 

That the forward and backward in time motions continually occur can also 
be seen by constructing the forward and backward in time velocities 
\be
v_{\pm }={\partial {\cal H}_{Brownian}\over \partial p_{\pm }}
=\pm \, \frac{1}{M}\le( p_\pm \mp \frac{R}{2}\, x_\mp \ri). \lab{(9)}
\ee 
These velocities do not commute 
\be
[v_+,v_-]=i\hbar \,{R\over M^2},  \lab{(10)}
\ee
and it is thereby impossible to fix the velocities forward and backward 
in time as being identical. Note the similarity between Eq.(\ref{(10)}) 
and the 
usual commutation relations for the quantum velocities 
${\bf v}=({\bf p}-(e{\bf A}/c))/M$ of a charged particle moving in a 
magnetic field ${\bf B}$; i.e. $[v_1,v_2]=(i\hbar eB_3/M^2c)$. Just as 
the magnetic field ${\bf B}$ induces a Aharonov-Bohm phase interference 
for the charged particle, the Brownian motion friction coefficient $R$ 
induces a closely analogous phase interference between forward and 
backward motion which expresses itself as mechanical damping. This part 
of the picture is discussed in Sec.4. Sec.5 is devoted to concluding
remarks.

\section{Two Slit Diffraction}

Shown in Fig.1 is a picture of a two slit experiment. What is 
required to derive the diffraction pattern is knowledge of the wave 
function $\psi_0(x)$ of the particle at time zero when it ``passes through 
the slits'', or equivalently the density matrix 
\be
(x_+|\rho_0|x_-)=\psi^*_0 (x_-)\psi_0 (x_+). \lab{(11)}
\ee
At a latter time $t$ we wish to find the probability density for the 
electron to be found at position $x$ at the detector screen,
\be
P(x,t)=(x|\rho (t)|x)=\psi^* (x,t)\psi (x,t). \lab{(12)}
\ee 
The solution to the free particle Schr\"odinger equation is
\be
\psi (x,t)=\Big({M\over 2\pi\hbar it} \Big)^{1/2} \int_{-\infty}^{\infty}
\exp\le[\frac{i}{\hbar } A(x-x^\prime,t)  \ri]\psi_0 (x^\prime )
\;dx^\prime ,\lab{(13a)}
\ee
where 
\be
A(x-x^\prime,t)={M(x-x^\prime )^2\over 2t} \lab{(13b)}
\ee
is the Hamilton-Jacobi action for a classical free particle to move from 
$x^\prime $ to $x$ in a time $t$. Eqs.(\ref{(11)})-(\ref{(13b)}) imply that 
\be
P(x,t)={M\over 2\pi\hbar t} 
\int_{-\infty}^{\infty}\int_{-\infty}^{\infty}
\exp\le[iM{(x-x_+)^2-(x-x_-)^2 \over 2\hbar t} \ri]
(x_+|\rho_0|x_-)\;dx_+dx_-. \lab{(14)}
\ee

The crucial point is that the density matrix $(x_+|\rho_0|x_-)$ when 
the electron ``passes through the slits'', depends non-trivially on the 
difference $(x_+-x_-)$ between the forward in time and backward in 
time coordinates. Were $x_+$ and $x_-$ always the same, then Eq.(\ref{(14)}) 
would imply that $P(x,t)$ not oscillate in $x$, i.e. there would 
not be the usual quantum diffraction. What is required for quantum 
interference in Eq.(\ref{(14)}) (cf. also Eq.(\ref{(13b)}) )
is that the forward in time action 
$A(x-x_+,t)$ differs from the backward in time action $A(x-x_-,t)$ 
for the phase interference to appear in the final probability 
density $P(x,t)$. 

\begin{figure}[t]
\setlength{\unitlength}{1mm}
\vspace*{80mm}
\includegraphics{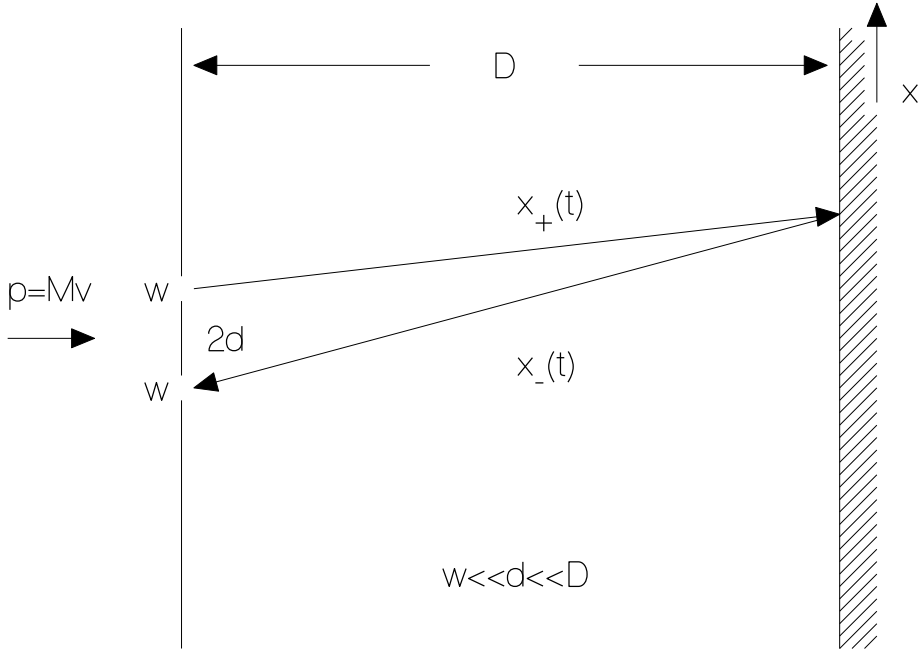}
\caption{Two slit experiment.}
\end{figure} 


For the usual quantum diffraction limit (see Fig.1)
\be
w \ll d \ll D, \lab{(15)}    
\ee
the diffraction pattern is adequately described by $|x|\gg|x_\pm|$; 
i.e. 
\be
P(x,t)\approx {M\over 2\pi\hbar t} 
\int_{-\infty}^{\infty}\int_{-\infty}^{\infty}
\exp\le[-i{M\,x\,(x_+-x_-)\over \hbar t} \ri]
(x_+|\rho_0|x_-)\;dx_+dx_-. \lab{(16)}
\ee
For the initial wave function we write 
\be
\psi_0(x)={1\over \sqrt{2}}\Big[\phi (x-d)+\phi (x+d)\Big], \lab{(17)}
\ee
where 
\be
\phi(x)={1\over \sqrt{w}} \quad if \quad 
|x|\leq \frac{w}{2} \quad and \quad 0 \quad 
otherwise. \lab{(18)} 
\ee
Eqs.(\ref{(11)}) and (\ref{(17)}) imply that
\bea
(x_+|\rho_0|x_-)&=&{1\over 2}\Big\{\phi (x_+-d)\phi (x_--d)+ 
\\ 
&& +
\phi (x_++d)\phi (x_-+d)+\phi (x_+-d)\phi (x_-+d)
+\phi (x_++d)\phi (x_--d)\Big\}. \lab{(19)}
\eea
The four terms in Eq.(\ref{(19)}) describe, 
respectively, the electron going 
forward and backward in time through slit 1, forward and backward 
in time through slit  2, forward in time through slit 1 and backward 
in time through slit 2, and forward in time through slit 2 and backward 
in time through slit 1.

The integral in Eq.(\ref{(16)}) is elementary and yields 
\be
P(x,t)\approx {4 \hbar \ t \over \pi M w \ x^2} \; 
\cos^2\Big({Md\ x \over \hbar\ t} \Big)
\sin^2\Big({Mw \ x \over 2\hbar \ t }\Big). \lab{(20)}
\ee 
Defining 
\be
K={Mvd \over \hbar D},\ \ \beta={w\over d},
\ee
where $v=D/t$ is the velocity of the incident electron, Eq.(\ref{(20)}) 
reads 
\be
P(x,D)\approx {4\over \pi \beta K x^2} \;
\cos^2(Kx)
\sin^2(\beta Kx). \lab{(21)}
\ee
This conventional diffraction result is plotted in Fig.2.

\begin{figure}[t]
\setlength{\unitlength}{1mm}
\vspace*{80mm}
\includegraphics{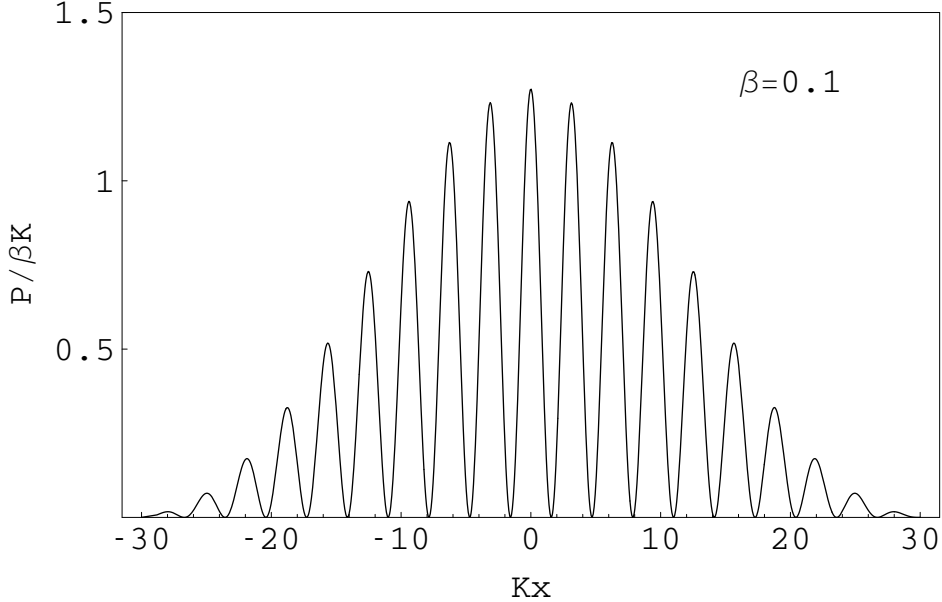}
\caption{Two slit interference pattern.}
\end{figure} 


\section{Quantum Mechanics with Dissipation}

The need to double the degrees of freedom of a Brownian motion particle 
is implicit even in the classical theory. Recall that in the classical 
Brownian theory one employs the equation of motion 
\be
M\ddot{x}(t)+R\dot{x}(t)=f(t), \lab{(22)}
\ee 
where $f(t)$ is a random (Gaussian distributed) force obeying 
\be
<f(t)f(t^\prime )>_{noise}=2\,R\,k_BT\; \delta (t-t^\prime). \lab{(23)}
\ee
To enforce Eq.(\ref{(22)}) one can employ a delta functional classical 
constraint representation as a functional integral
\be
\delta[M\ddot{x}+R\dot{x}-f]=
\int {\cal D}y \; \exp\le[{i\over \hbar}\,
\int dt \; y\{f-M\ddot{x}-R\dot{x}\}\ri]. \lab{(24)}
\ee
Note in Eq.(\ref{(24)}) that one needs a constant $\hbar $ with dimensions of 
action which from purely classical considerations cannot be fixed in 
magnitude. From the viewpoint of quantum mechanics, we know how to fix 
the magnitude. (Exactly the same situation prevails in the purely classical 
statistical mechanics of Gibbs. The dimensionless phase space volume 
is $\Pi_k (dp_kdq_k/2\pi \hbar )$ and the precise value to be chosen for 
the action quantum $2\pi \hbar $ was evident only after quantum theory.)

Integration by parts in the time integral of Eq.(\ref{(24)}), 
and averaging over 
the fluctuating force $f$ yields 
\be
<\delta[M\ddot{x}+R\dot{x}-f]>_{noise}=
\int {\cal D}y <\exp\le[{i\over \hbar}
\int dt \;{\cal L}_f(\dot{x},\dot{y},x,y)\ri]>_{noise}, \lab{(25)}
\ee
where 
\be
{\cal L}_f(\dot{x},\dot{y},x,y)=M\dot{x}\dot{y}+
{R\over 2}(x\dot{y}-y\dot{x})+fy. \lab{(26)}
\ee
At the classical level, the constraint condition introduced a new 
coordinate $y$, and from a Lagrangian viewpoint
%

\be
{d\over dt}{\partial {\cal L}_f\over \partial \dot{y}}=
{\partial {\cal L}_f\over \partial y} \quad ; \ \ \ 
{d\over dt}{\partial {\cal L}_f\over \partial \dot{x}}=
{\partial {\cal L}_f\over \partial x},  
\ee
i.e.
\be
M\ddot{x}+R\dot{x}=f \quad ;\ \ \ 
M\ddot{y}-R\dot{y}=0. \lab{(27)}
\ee
It is in fact true that the Lagrangian system Eqs.(\ref{(26)})-(\ref{(27)})  
were discovered from a completely classical viewpoint \cite{BA}. In the 
$x$ coordinate there is damping, but in the $y$ coordinate there is 
amplification.

Although the Lagrangian Eq.(\ref{(26)}) was not here motivated by quantum 
mechanics, it is a simple matter to make contact with the theory 
of a quantum Brownian particle moving in a classical fluid using 
the transformation \cite{SWV} 
\be
x_{\pm}=x\pm {y \over 2}. \lab{(28)}
\ee
In this case, after averaging over the random force using 
\be
<\exp\le[{i\over \hbar}\int dt \; y(t)f(t)\ri]>_{noise}=
\exp\le[-{k_BTR\over \hbar^2}\int dt \; y(t)^2\ri], \lab{(29)}
\ee
one finds
\be
<\exp\le[{i\over \hbar}\int dt \;{\cal L}_f(\dot{x},\dot{y},x,y)\ri]>_{noise}
=\exp\le[{i\over \hbar}\int dt \;{\cal L}(\dot{x}_+,\dot{x}_-,x_+,x_-)\ri], 
\lab{(30)}
\ee
with a complex Lagrangian
\be
{\cal L}(\dot{x}_+,\dot{x}_-,x_+,x_-)={M\over 2}(\dot{x}_+^2-\dot{x}_-^2)
+{R\over 2}(\dot{x}_+x_- - \dot{x}_-x_+)\,+ \,i
{k_BTR\over \hbar}(x_+-x_-)^2, \lab{(31)}
\ee
In evaluating Eq.(\ref{(29)}) 
we employed Eq.(\ref{(23)}) and the Gaussian nature 
of the random force. Considered as a statistical probability in the 
coordinate $y$, Eq.(\ref{(23)}) 
represents a Gaussian process with a correlation 
function given in Eq.(\ref{(8)}). 
Employing the Lagrangian in Eq.(\ref{(31)}) in a path 
integral formulation for the density matrix \cite{SW, FEV},   
\be
(x_+|\rho (t)|x_-)=
\int_{-\infty}^\infty \int_{-\infty}^\infty 
K(x_+,x_+^\prime ,x_-,x_-^\prime ,t)\,
(x_+^\prime|\rho_0|x_-^\prime) \;dx_+^\prime dx_-^\prime , \lab{(32a)}
\ee
and
\be
K(x_+,x_+^\prime ,x_-,x_-^\prime ,t)=
\int_{x_+^\prime}^{x_+} {\cal D}x_+(t^\prime) \int_{x_-^\prime}^{x_-} 
{\cal D}x_-(t^\prime)\ \exp\le[{i\over \hbar }\int_0^t 
{\cal L}^\prime dt^\prime \ri], \lab{(32b)} 
\ee
yields the equation of motion (\ref{(7c)}). 

Note, from Eqs.(\ref{(7a)})-(\ref{(7c)}) the normalization integral 
\be
N(t)=Tr\rho (t)=\int_{-\infty}^\infty \int_{-\infty}^\infty 
\delta (x_+-x_-)\;(x_+|\rho (t)|x_-)\;dx_+ dx_- \lab{(34a)}
\ee
obeys 
\be
\dot{N}(t)=- {R\over 2M}N(t)\ ,\ \  \ \  N(t)=N(0)e^{-\frac{R}{2M}t}. 
\lab{(35b)}
\ee
The decay of the normalization is a consequence of the customary 
procedure of integrating out an infinite number of thermal Brownian motion 
bath coordinates in statistical mechanics. This process gives rise 
to an effective Hamiltonian ${\cal H}_{Brownian}$ which even in the limit 
$T\to 0$ is not self adjoint; i.e. with  
\be
{\cal H}_0=\lim_{T\to 0}\ {\cal H}_{Brownian}, \lab{(36)}
\ee 
the eigenvalues of ${\cal H}_0$ are complex \cite{FE, QD}. These complex 
eigenvalues lead to the (temperature independent) decay of $N(t)$. To 
keep the probability ``normalized'' one merely uses the average 
$<A>=(tr(\rho A)/tr\rho)$.
 
At zero temperature, the equation of motion for the density matrix 
is given by  
\be
i\hbar {\partial (x_+|\rho (t)|x_-) \over \partial t}=
{\cal H}_0\ (x_+|\rho (t)|x_-), \lab{(37a)}
\ee
\be
{\cal H}_0\ =
\frac{1}{2M}\le(p_+ -  \frac{R}{2} \, x_-\ri)^2 -
\frac{1}{2M}\le(p_- + \frac{R}{2} \, x_+\ri)^2. \lab{(37b)}
\ee
The solution to Eq.(\ref{(37a)}) has the form 
\be
(x_+|\rho (t)|x_-)=
\int_{-\infty}^\infty \int_{-\infty}^\infty 
K_0 (x_+,x_+^\prime ,x_-,x_-^\prime ,t)\,
(x_+^\prime|\rho_0|x_-^\prime) \; dx_+^\prime dx_-^\prime , \lab{(38)}
\ee
where 
\be
K_0 (x_+,x_+^\prime ,x_-,x_-^\prime ,t)=
e^{-i{\cal H}_0t/\hbar}\delta (x_+-x_+^\prime )\delta (x_--x_-^\prime ), 
\lab{(39)}
\ee
or, in path integral form 
\be
K_0 (x_+,x_+^\prime ,x_-,x_-^\prime ,t)=
\int_{x_+^\prime}^{x_+} {\cal D}x_+(s) \int_{x_-^\prime}^{x_-} 
{\cal D}x_-(s)\ \exp\le[{i\over \hbar }\int_0^t 
{\cal L}_0(s) ds\ri], \lab{(40)} 
\ee
where 
\be
{\cal L}_0(s)
={M\over 2}\Big[\dot{x}_+^2(s)-\dot{x}_-^2(s)\Big]
+{R\over 2}\Big[ \dot{x}_+(s)x_-(s)-\dot{x}_-(s)x_+(s)\Big]. \lab{(41)}
\ee
From Eqs.(\ref{(40)}), (\ref{(41)}), 
a translation of the coordinates by a constant 
$(x_+,x_-)\to (x_++a_+,x_-+a_-)$ yields
\bea
&&K_0 (x_++a_+,x_+^\prime +a_+,x_-+a_-,x_-^\prime +a_-,t)= \hspace{8cm}
\\
&& \hspace{4cm}
\exp\le[{iR\over 2\hbar}\Big(a_-(x_+-x_+^\prime)- 
a_+(x_--x_-^\prime)\Big)\ri]
K_0 (x_+,x_+^\prime ,x_-,x_-^\prime ,t). \lab{(42)}
\eea
From Eq.(\ref{(42)}), 
\be
K_0 (x_+,x_+^\prime ,x_-,x_-^\prime ,t)=
e^{i\Phi (x_+,x_-,x_+^\prime ,x_-^\prime )}
{\cal F}_0 (x_+-x_+^\prime ,x_--x_-^\prime ,t), \lab{(43a)}
\ee
where 
\be
\Phi (x_+,x_-,x_+^\prime ,x_-^\prime )=
{R\over 2\hbar }(x_+x_-^\prime -x_-x_+^\prime). 
\lab{(43b)}
\ee
From 
\be
i\hbar {\partial  K_0 (x_+,x_+^\prime ,x_-,x_-^\prime ,t) 
\over \partial t}={\cal H}_0
\ K_0 (x_+,x_+^\prime ,x_-,x_-^\prime ,t), \lab{(44)}
\ee
and Eq.(\ref{(43a)}) one finds 
\be
i \hbar {\partial {\cal F}_0 (x_+,x_-,t)\over \partial t}=
\le[\frac{1}{2M}\le(p_+- \frac{R}{2} \, x_-\ri)^2 \,-
\,\frac{1}{2M}\le(p_- + \frac{R}{2}
\, x_+ \ri)^2 \ri]{\cal F}_0 (x_+,x_-,t). \lab{(46)}
\ee
With 
\be
\gamma =\frac{R}{2M}, \lab{(47a)}
\ee
the solution of Eq.(\ref{(46)}) is given by 
\be
{\cal F}_0 (x_+,x_-,t)=
{M\gamma \over 2\pi \hbar\ \sinh(\gamma t)}
\exp\le[\frac{i}{2\hbar} M\,\gamma \ \coth(\gamma t)(x_+^2-x_-^2)\ri].
\lab{(47b)}
\ee

\vspace{.5cm}

\section{Phase Coherence and Dissipative Flux}

The above results may be applied to the two slit diffraction 
problem as in Eq.(\ref{(14)}). 
The general result is that the probability 
density in $x$ is given by 
\be
P(x,t)=\int_{-\infty}^{\infty}\int_{-\infty}^{\infty}
K(x,x_+,x,x_-,t)\,(x_+|\rho_0|x_-)\;dx_+dx_-. \lab{(48)}
\ee 
In the regime of Eq.(\ref{(16)}) we then obtain from 
Eqs.(\ref{(43a)}),(\ref{(43b)}) and (\ref{(47a)})-(\ref{(48)})  
and the renormalized time $\tau $ 
\be
\gamma \tau =e^{-\gamma t}\sinh(\gamma t), \lab{(49a)}
\ee 
\be
e^{\gamma t}P_0(x,t)\approx 
{M\over 2\pi\hbar \tau} 
\int_{-\infty}^{\infty}\int_{-\infty}^{\infty}
\exp\le[-\frac{iM}{\hbar \tau} \,x\, (x_+-x_-) \ri]
(x_+|\rho_0|x_-)\;dx_+dx_-. \lab{(49b)}
\ee 

Comparing Eq.(\ref{(16)}) to Eq.(\ref{(49b)}) 
one finds the following remarkable 
result: For a particle in a bath which induces a damping $\gamma =(R/2M)$ 
at zero temperature, the slit diffraction patterns for 
the frictional case can be obtained from those of the zero friction 
case. All that is required is to rescale the effective time 
according to Eq.(\ref{(49a)}).  

The probability density (at zero temperature) to find a particle in 
the interval $dx$ is proportional to 
\be
P(x,t)=
{M\gamma \over 2\pi \hbar \sinh(\gamma t)}
\int_{-\infty}^{\infty}\int_{-\infty}^{\infty}
\exp \le[i\, \phi (x,x_+-x_-)+ i \,{M\gamma (x_+^2-x_-^2)
\over 2\hbar \tanh(\gamma t) }\ri]
(x_+|\rho_0|x_-)\;dx_+dx_-, \lab{(50a)}
\ee
where  
\be
\phi (x,x_+-x_-)=-{M \gamma \,x (x_+-x_-)\over \hbar}
=-{R \,x (x_+-x_-)\over 2\hbar}. \lab{(50b)} 
\ee
Eqs.(\ref{(50a)}), (\ref{(50b)}) follow from Eqs.(\ref{(43a)}),
 (\ref{(43b)})
and (\ref{(47a)})-(\ref{(48)}). 

The phase in Eq.(\ref{(43b)}), i.e. $\Phi (x_+,x_-,x_+^\prime ,x_-^\prime )=
(R/ 2\hbar)(x_+x_-^\prime - x_-x_+^\prime )$ represents a ``dissipative 
flux'' $2 \hbar \Phi =R\ Area$. With ${\bf X}=(x_+,x_-)$ and 
${\bf X}^\prime =(x_+^\prime ,x_-^\prime) $ as vectors in a plane,  
$Area={\bf N\cdot }({\bf X \times X}^\prime )$,  
where ${\bf N}$ is the vector normal to the plane. The phase 
$
\phi (x,x_+^\prime -x_-^\prime )=\Phi (x_+=x,x_-=x,x_+^\prime,x_-^\prime)
$ 
in Eqs. (\ref{(50a)}) and (\ref{(50b)}) is of the dissipative flux type. 
 
Note the similarity between Eq.(\ref{(37b)}) 
and the Hamiltonian for a particle 
in the $x-y$ plane with a magnetic field in the z-direction; i.e. 
$H=({\bf p}-e\,{\bf B\times r}/2c)^2/2M$. For the magnetic field case, 
the flux is $B\times Area $ while for the closely analogous case of 
${\cal H}_0$ the flux is $R\times Area$.  
Magnetic flux induces Aharonov-Bohm phase interference 
for the charged particle. Dissipative flux
 yields an analogous phase interference between forward and 
backward in time motion as expressed by mechanical damping. 
The similarity is most easily appreciated in the path integral 
formulation as in Eqs.(\ref{(31)}) and (\ref{(32a)}), (\ref{(32b)}). 
The resistive part of the action is
\be
{\cal S}_R={R\over 2}\int (x_-\dot{x}_+ - x_+\dot{x}_- ) dt=
{R\over 2}\int (x_-dx_+ - x_+dx_-).
\lab{(51)}
\ee 
To see the phase interference for two different paths $P_1$ and 
$P_2$ in the $(x_+,x_-)$ plane with the same endpoints,  
one should compute (with $\int_{P_1}-\int_{P_2}=\oint$), 
\be
{{\cal S}_{R}(\sma{interference})\over \hbar }={R\over 2\hbar }
\oint (x_-dx_+-x_+dx_-)={R\, \Sigma (P_1,P_2)\over \hbar },
\lab{(52)}
\ee
where $\Sigma (P_1,P_2)$ is the oriented area between the two paths $P_1$ 
and $P_2$. Such phase interference $\exp\le[i R \,\Sigma /\hbar\ri]$ enters 
into the path integral formulation of the problem in Eq.(\ref{(40)}). 
The condition of constructive phase interference is thereby  
$R\,\Sigma =2\pi n \hbar $ where $n=0,\pm 1,\pm 2,...$ is a  
quantization integer.

\section{Conclusions}

In the conventional textbook description of quantum mechanics, 
one considers that there is but one coordinate $x$ 
(or more generally one set of coordinates) which describes a 
physical system. As shown by Schwinger (in his seminal work on 
Brownian motion \cite{SW}), in quantum mechanical theory it is 
often more natural to consider doubling the system coordinates, 
in our case from one coordinate $x(t)$ describing motion in time 
to two coordinates, say $x_+(t)$ going forward in time 
and $x_-(t)$ going backward in time. 

In this picture, a system acts in a classical fashion if 
the two paths can be identified, i.e. 
$x_{classical}(t)\equiv x_{+\ classical}(t)\equiv x_{-\ classical}(t)$. 
When the system moves so that the forward in time and backward in time 
motions are (at the same time) unequal $x_+(t)\ne x_-(t)$,
then the system is behaving in a quantum mechanical fashion and 
exhibits interference patterns in measured position probability densities.
Of course when $x$ is actually measured there is only one {\it classical} 
$x=x_+=x_-$. 

It is only when you do not look at a coordinate, e.g. do not look at which 
slit the electron may have passed, that the quantum picture is valid 
$x_+\ne x_-$. In this fascinating regime in which coordinate doubling 
and path splitting takes place, we are all under the dictates of Bohr who 
finally warns us not to ask what the quantum system is really doing.  
When the system is quantum mechanical just add up the amplitudes and 
absolute square them. Ask nothing more.

In this work we have concentrated on the low temperature limit, which 
means $T \ll T_\gamma $ where 
\be
k_BT_\gamma=\hbar \gamma={\hbar R\over 2M}. \lab{(53)}
\ee
In the high temperature regime $T \gg T_\gamma $, the thermal bath 
motion suppresses the probability for $x_+\ne x_-$ via the thermal term 
$(k_BTR /\hbar)(x_+-x_-)^2$ in Eq.(\ref{(7a)}). In terms of the diffusion 
coefficient in Eqs.(\ref{(1)}) and (\ref{(53)}), i.e. 
\be
D={T\over T_\gamma }\Big({\hbar \over 2M}\Big), \lab{(54)}
\ee
the condition for classical Brownian motion for high mass particles 
is that $D \gg (\hbar /2M)$, and the condition for quantum interference
with low mass particles is that $D \ll (\hbar /2M)$. For a single atom 
in a fluid at room temperature it is typically true that 
$D\sim (\hbar /2M)$, equivalently $T\sim T_\gamma $ so that quantum 
mechanics plays an important but perhaps not dominant role in the 
Brownian motion. For large particles (in, say, colloidal systems) 
classical Brownian motion would appear to dominate the motion.  
It is interesting to note that in the formulation of quantum mechanics 
known as stochastic quantization, $(\hbar /2M)$ plays the role of a 
diffusion coefficient of a sort defined by Nelson \cite{NE} which also 
distinguishes forward and backwards in time splitting. In such a 
formulation the distinction between low temperature quantum motions 
and high temperature classical motions would become the distinction 
between Nelson diffusion and Einstein diffusion.

It is remarkable that, although in different contexts and in  different 
view point frameworks, coordinate doubling has also entered into the
canonical quantization of finite temperature field theoretical systems
\cite{TFD} as well as other dissipative systems \cite{FE, QD} and it 
appears to be intimately related to the algebraic properties of the
theory \cite{DM, KH}. 

Finally, we note that the "negative" kinematic term in the Lagrangian 
(38) also appears in two-dimensional gravity models leading to (at least) 
two different strategies in the quantization method \cite{JA}: the 
Schr\"odinger representation approach, where no negative norm appears, and 
the string/conformal field theory approach where negative norm states 
arise as in Gupta-Bleurer electrodynamics. It appears to be an 
interesting question to ask about any deeper connection between the 
$(x_+,x_-)$ Schwinger formalism and the subtelties of low 
dimensional gravity theory.

We hope that the views discussed in this work have clarified the 
nature of coordinate doubling framework.

\vspace{0.7cm}

\noindent{\bf Acknowledgments}

This work has been partially supported by DOE in USA, by INFN in Italy and
by EU Contract ERB CHRX CT940423.


\newpage

\begin{thebibliography}{9999}


\bibitem{TO} A. Tonomura, J. Endo, T. Matsuda, T.Kawasaki and H. Exawa,
{\it Amer. J. Phys.} {\bf 57} (1989), 117

\bibitem{SW} J. Schwinger, {\it J. Math. Phys. }{\bf 2} (1961), 407

\bibitem{DO} V.V. Dodonov, O.V. Man'ko and V.I. Man'ko, {\it J. of
Russian Laser Research} {\bf 16} (1995), 1

\bibitem{SWV} Y.N. Srivastava, G. Vitiello and A. Widom, {\it Ann. 
Phys. (N.Y.) }{\bf 238} (1995), 200

\bibitem{CS} M. Blasone, E. Graziano, O.K. Pashaev and G. Vitiello,
{\it Ann. Phys. (N.Y.)} {\bf 252} (1996), 115 

\bibitem{TS} Y. Tsue, A. Kuriyama and M. Yamamura,  {\it Progr. Theor. Physics}
{\bf 91} (1994), 469

\bibitem{BA} H. Bateman, {\it Phys. Rev.} {\bf 38} (1931), 815

\bibitem{FEV} R.P. Feynman and F.L. Vernon, {\it Ann. Phys. (N.Y.)} {\bf
24} (1963), 118

\bibitem{FE} H. Feshbach and Y. Tikochinsky, {\it Trans. New York Acad. 
Sci. (Ser.II)} {\bf 38} (1977), 44

\bibitem{QD} E. Celeghini, M. Rasetti and G. Vitiello, {\it Ann. 
Phys. (N.Y.) }{\bf 215} (1992), 156

\bibitem{NE} E. Nelson, {\it Quantum Fluctuations} (Princeton 
University Press, Princeton, 1985)

\bibitem{TFD} Y. Takahashi and H. Umezawa, {\it Collective Phenomena }
{\bf 2} (1975), 55; U. Umezawa, M. Matsumoto and M. Tachiki, 
{\it Thermo Field Dynamics and Condensed States} 
(North-Holland, Amsterdam, 1982)

\bibitem{DM} S. De Martino, S. De Siena and G. Vitiello, {\it Int. J. Mod.
Phys.} {\bf B10} (1996), 1615

\bibitem{KH} A.E. Santana and F.C. Khanna, {\it Phys. Lett.} 
{\bf A203} (1995), 68


\bibitem{JA} R. Jackiw, {\it Two lectures on two-dimensional gravity},
gr-qc/9511048; D. Cangemi, R. Jackiw and B. Zwiebach, {\it Ann. Phys. 
(N.Y.)} {\bf 245} (1996), 408

\end{thebibliography}
\end{document}